\documentclass[prm,preprint,amsmath,amssymb,superscriptaddress]{revtex4-1}
\usepackage{graphicx}
\usepackage[utf8]{inputenc}
\usepackage{amssymb}
\usepackage{xcolor}
\usepackage{upgreek}
\usepackage{units}
\usepackage{float}
\usepackage{amsmath}
\usepackage{url}
\usepackage{epsfig}
\usepackage{bm}
\usepackage{xspace} 
\usepackage{hyperref}
\usepackage{lineno}
\usepackage{xr-hyper}
\hypersetup{
  colorlinks  = true, 
  urlcolor   = blue,
  linkcolor  = red, 
  citecolor  = blue,
  filecolor = cyan % color for links to external docs (xr-hyper), SI references
}

\begin{document}

\title{Direct demonstration of time-reversal-symmetry-breaking spin injection from a compensated magnet}
\author{Jone Mencos}
\affiliation {CIC nanoGUNE BRTA, 20018 Donostia-San Sebastián, Basque Country, Spain.}
\affiliation{Departamento de Polímeros y Materiales Avanzados: Física, Química y Tecnología, University of the Basque Country (UPV/EHU), 20018 Donostia-San Sebastián, Basque Country,
Spain.}
\author{Antonin Badura}
\affiliation{Institute of Physics, Czech Academy of Sciences, Prague, Czechia.}
\affiliation{Faculty of Mathematics and Physics, Charles University, Prague, Czechia.}
\author{Eoin Dolan}
\affiliation {CIC nanoGUNE BRTA, 20018 Donostia-San Sebastián, Basque Country, Spain.}
\affiliation{Departamento de Polímeros y Materiales Avanzados: Física, Química y Tecnología, University of the Basque Country (UPV/EHU), 20018 Donostia-San Sebastián, Basque Country,
Spain.}
\author{Sebastian Beckert}
\affiliation{Dresden Center for Nanoanalysis, cfaed, TUD University of Technology Dresden, 01069 Dresden, Germany}
\author{Rafael González-Hernández}
\affiliation{Departamento de Física, Universidad del Norte, Barranquilla, Colombia.}
\affiliation{Institut für Physik, Johannes Gutenberg Universität Mainz, Mainz, Germany.}
\author{\textcolor{black}{Nicolás Sigales}}
\affiliation{\textcolor{black}{Instituto de Física, Facultad de Ciencias, UdelaR, Montevideo, Uruguay}}
\affiliation{\textcolor{black}{Grupo disciplinar Física, Departamento Suelos y Aguas, Facultad de Agronomía, UdelaR, Montevideo, Uruguay}}
\author{\textcolor{black}{Tim Kokkeler}}
\affiliation{\textcolor{black}{Department of Physics and Nanoscience Center, University of Jyvaskyla,
P.O. Box 35 (YFL), FI-40014 University of Jyvaskyla, Finland}}
\author{Ismaïla Kounta}
\affiliation{Aix-Marseille Univ., CNRS, CINaM, Marseille, France.}
\author{Matthieu Petit}
\affiliation{Aix-Marseille Univ., CNRS, CINaM, Marseille, France.}
\author{Charles Guillemard}
\affiliation{Aix-Marseille Univ., CNRS, CINaM, Marseille, France.}
\author{Anna Birk Hellenes}
\affiliation{Institute of Physics, Czech Academy of Sciences, Prague, Czechia.}
\affiliation{Institut für Physik, Johannes Gutenberg Universität Mainz, Mainz, Germany.}
\author{Warlley Campos}
\affiliation{Institut für Physik, Johannes Gutenberg Universität Mainz, Mainz, Germany.}
\affiliation{Max Planck Institute for the Physics of Complex Systems, N\"othnitzer Str.\ 38, 01187 Dresden, Germany}
\author{Javier Rial}
\affiliation{Univ. Grenoble Alpes, CNRS, CEA, Grenoble INP, IRIG-SPINTEC, F-38000 Grenoble, France.}
\author{Dominik Kriegner}
\affiliation{Institute of Physics, Czech Academy of Sciences, Prague, Czechia.}
\author{Vincent Baltz}
\affiliation{Univ. Grenoble Alpes, CNRS, CEA, Grenoble INP, IRIG-SPINTEC, F-38000 Grenoble, France.}
\author{Luis E. Hueso}
\affiliation{CIC nanoGUNE BRTA, 20018 Donostia-San Sebastián, Basque Country, Spain.}
\affiliation{IKERBASQUE, Basque Foundation for Science, 48009 Bilbao, Basque Country, Spain.}
\author{Jairo Sinova}
\affiliation{Institut für Physik, Johannes Gutenberg Universität Mainz, Mainz, Germany.}
\author{\textcolor{black}{F. Sebastian Bergeret}}
\affiliation{\textcolor{black}{Centro de Física de Materiales (CFM-MPC) Centro Mixto CSIC-UPV/EHU, 20018 Donostia-San Sebastián, Basque Country, Spain}}
\affiliation{\textcolor{black}{Donostia International Physics Center (DIPC), 20018 Donostia-San Sebastián, Basque Country, Spain}}
\author{Olena Gomonay}
\affiliation{Institut für Physik, Johannes Gutenberg Universität Mainz, Mainz, Germany.}
\author{Tomáš Jungwirth}
\affiliation{Institute of Physics, Czech Academy of Sciences, Prague, Czechia.}
\affiliation{School of Physics and Astronomy, University of Nottingham, Nottingham, United Kingdom.}
\author{Libor Šmejkal}
\affiliation{Max Planck Institute for the Physics of Complex Systems, N\"othnitzer Str.\ 38, 01187 Dresden, Germany}
\affiliation{Max Planck Institute for Chemical Physics of Solids, N\"othnitzer Str.\ 40, 01187 Dresden, Germany}
\affiliation{Institute of Physics, Czech Academy of Sciences, Prague, Czechia.}
\author{Lisa Michez}
\affiliation{Aix-Marseille Univ., CNRS, CINaM, Marseille, France.}
\author{Helena Reichlova}
\email{reichlh@fzu.cz}
\affiliation{Institute of Physics, Czech Academy of Sciences, Prague, Czechia.}
\author{Fèlix Casanova}
\email{f.casanova@nanogune.eu}
\affiliation{CIC nanoGUNE BRTA, 20018 Donostia-San Sebastián, Basque Country, Spain.}
\affiliation{IKERBASQUE, Basque Foundation for Science, 48009 Bilbao, Basque Country, Spain.}

% \date{\today}
\maketitle
%\linenumbers
%ABSTRACT

{\bf 
The injection, propagation and detection of spin currents are essential physical processes in spintronics. So far, the separation of charge and spin currents was facilitated by a direct exploitation of the electrical spin injection from a ferromagnet \cite{johnson_interfacial_1985, jedema_electrical_2001,jedema_spin_2003,tombros_electronic_2007,lou_electrical_2007,casanova_control_2009,hamaya_estimation_2012,zahnd_spin_2018,fukuma_giant_2011,Kaiser2024} or the injection by a relativistic spin Hall effect \cite{sagasta_tuning_2016, morota_indication_2011, laczkowski_large_2017,omori_relation_2019, m_cosset-cheneau_electrical_2022}. The devices employed are lateral spin valves comprising spatially well-separated injection and detection electrodes, connected by a spin-propagation channel. The time-reversal ($\cal T$) symmetry-breaking ferromagnetic spin injection is realized in an experimental geometry with an electrical bias applied between the injection electrode and the channel, and is modeled by a conserved spin-polarized drift current generated along the applied bias \cite{johnson_interfacial_1985, jedema_electrical_2001}. In contrast, the spin injection by the $\cal T$-symmetric relativistic spin Hall mechanism is driven by an electrical bias applied across the injection electrode alone, and is modeled by a non-conserved spin current transverse to the applied bias \cite{Sinova2015}. In this work, we use a lateral spin valve with a Mn$_5$Si$_3$ injection electrode to directly demonstrate a $\cal T$-symmetry-breaking spin injection from a compensated magnet with a vanishing net magnetization. Specifically, the $\cal T$-symmetry breaking is demonstrated by the fact that switching between time-reversed states of the compensated magnet changes the detected spin signal. Moreover, the $\cal T$-symmetry-breaking nature of the spin injection is observed in both electrical biasing geometries while employing the same detection electrode. We show that this unconventional spin-injection phenomenology is consistent with different magnitudes and propagation angles of electrical currents in the spin-up and spin-down channel in a d-wave altermagnet. \cite{Gonzalez-Hernandez2021,smejkal_giant_2022,Smejkal2021a,Smejkal2022a,jungwirth2026altermagnetic} Here our symmetry analysis and first-principles calculations are based on the compensated collinear altermagnetic order which has provided a comprehensive microscopic interpretation of earlier structural, magnetic, and anomalous Hall and Nernst measurements in Mn$_5$Si$_3$ thin films \cite{Reichlova2024,kounta_competitive_2023,Badura2025}.
}

%INTRO

It is now theoretically and experimentally established that some compensated magnets with vanishing net magnetization can generate $\cal T$-symmetry-breaking responses that were traditionally considered the exclusive domain of ferromagnetic spintronics \cite{Smejkal2022AHEReview,Smejkal2022a,Nakatsuji2022,Rimmler2024,gonzalez_betancourt_spontaneous_2023,jungwirth2026altermagnetic}. \textcolor{black}{Analogously to ferromagnets, these responses enable the control and detection of switching between time-reversed states of a compensated magnet. At the same time, unlike in ferromagnets, they open the prospect of future spintronic technologies whose energy efficiency, speed, and scalability are not limited by magnetization \cite{Smejkal2022AHEReview,Smejkal2022a,Nakatsuji2022,Rimmler2024,Han2025,jungwirth2026altermagnetic}.} To enable the $\cal T$-symmetry-breaking  linear responses,  the compensated magnetic ordering on the crystal has to break symmetries combining $\cal T$ with translation and $\cal T$ with inversion. These broken symmetries can occur in some non-collinear compensated magnets, and are among the defining symmetry signatures of the collinear compensated altermagnetic phase. For the non-collinear magnetic orders, the resulting spin-polarized spectra tend to feature mixed spin-up and spin-down electronic states, reminiscent of a strong relativistic spin-orbit coupling \cite{Zelezny2017a,Smejkal2022AHEReview}. In contrast, the collinear altermagnets feature spin-up and spin-down transport channels which can be well separated and conserved \cite{Smejkal2021a,Smejkal2022a}, as in conventional ferromagnets used in current spintronic technologies \cite{dieny2020opportunities,incorvia2024spintronics,Worledge2022,IRDS2024}.

Research of $\cal T$-symmetry-breaking phenomena in compensated magnets has so far focused on local responses. These include the counterparts of the ferromagnetic anomalous Hall (AHE) and anomalous Nernst (ANE) effects, measured in a thin-film cross-bar structure of the studied magnet, or the giant (tunneling) magnetoresistance, detected in a vertical stack with magnetic electrodes separated by a nanoscale non-magnetic conductive (insulating) spacer \cite{Nakatsuji2015,gonzalez_betancourt_spontaneous_2023,Chen2023,Qin2023, ikhlas_large_2017, Badura2025}. \textcolor{black}{Unlike these local measurements, the non-local geometry of a lateral spin valve (LSV) directly probes spin-current generation and propagation away from the compensated magnet. Here, we integrate  a compensated magnet in a LSV to enable such a non-local detection scheme.}

\textcolor{black}{Figure 1 summarizes that the spin-injection phenomenology observed in the compensated magnet} is unparalleled among the previously explored ferromagnetic and relativistic spin-injection mechanisms. In the top row, we depict the two electrical biasing geometries that can be used for spin injection in LSVs. Spin injection in the first biasing geometry due to a spin-polarized longitudinal drift current (left cartoon) has been demonstrated in ferromagnetic materials \cite{johnson_interfacial_1985, jedema_electrical_2001,jedema_spin_2003,tombros_electronic_2007,lou_electrical_2007,casanova_control_2009,hamaya_estimation_2012,zahnd_spin_2018,fukuma_giant_2011,Kaiser2024}, where the breaking of $\cal{T}$-symmetry originates from the net magnetization. In the second biasing geometry (right cartoon), the spin injection relies on $\cal{T}$-symmetric relativistic spin–orbit coupling effects, namely the transverse spin Hall current, and has been observed using both ferromagnetic \cite{omori_relation_2019, m_cosset-cheneau_electrical_2022} and non-magnetic heavy-metal injection electrodes \cite{sagasta_tuning_2016, morota_indication_2011, laczkowski_large_2017}. The spin accumulation generated in a non-magnetic channel in both geometries diffuses away as a spin current and is non-locally detected as a voltage signal by a ferromagnetic electrode which depends on the relative orientation between its magnetization and the polarization direction of the diffusing spin current. \textcolor{black}{In our work, we directly observe a $\cal T$-symmetry-breaking spin injection from a compensated magnet by measuring the reversal of the non-local signal upon switching between time-reversed states of the compensated magnet. Remarkably, the same $\cal T$-symmetry-breaking spin-current generation is revealed in both electrical biasing geometries. Our results demonstrate that a spin current can be generated and controlled by a magnetic state with no net magnetization. This control by the magnetic state of a compensated magnet makes the observed phenomenon fundamentally distinct from the $\cal T$-symmetric relativistic spin Hall effect, while the absence of magnetization distinguishes it from conventional ferromagnetic spin injection.}

\begin{figure}[!ht]
    \centering
    \includegraphics[width=0.9\textwidth]{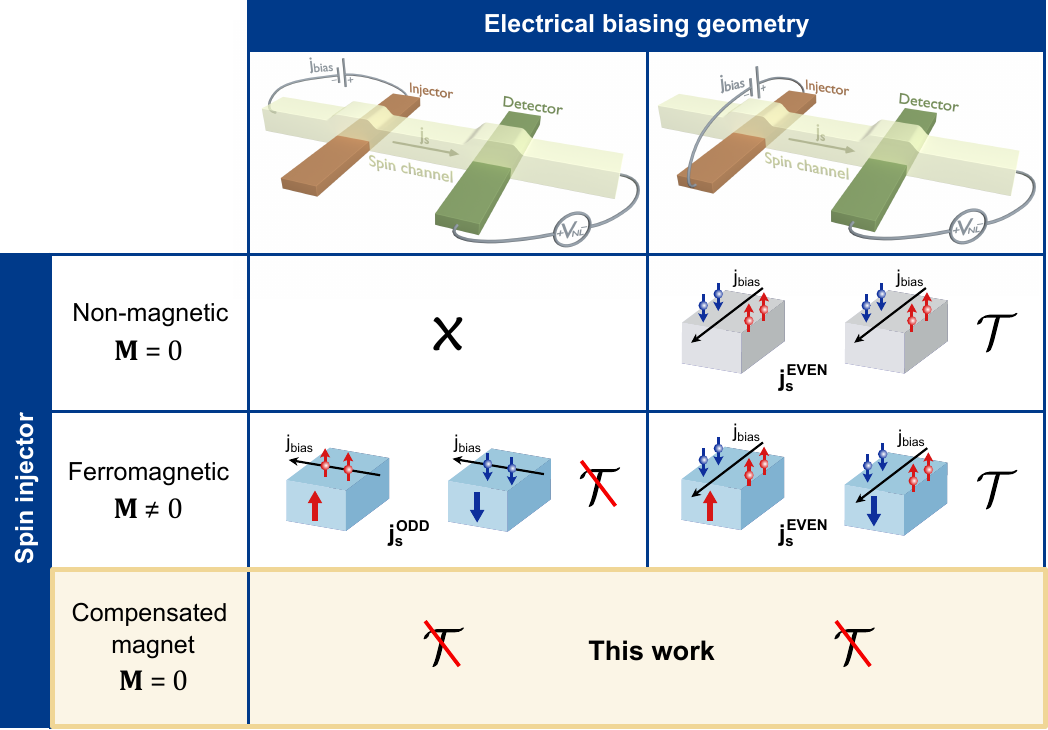}
	\caption{\textbf{Spin-injection mechanisms and direct detection in a lateral spin valve.} Top row: A LSV comprising injection and detection electrodes connected by a transverse channel. An applied electrical bias ($\textit{I}_\text{bias}$) injects spins into the channel, which then diffuse towards the detection electrode, generating a non-local voltage, $\textit{V}_\text{NL}$. The left and right schematics illustrate the first and the second electrical biasing geometries in which the respective spin-injection mechanisms have been previously demonstrated. Second row: $\cal{T}$-symmetric relativistic spin injection \textcolor{black}{of $\cal{T}$-even spin current ($\textit{j}_\text{S}^{EVEN}$)} from a non-magnetic electrode, arising from a transverse spin-Hall current, in the second biasing geometry. In the first biasing geometry, spin injection from a non-magnetic electrode is not observed. Third row: $\cal{T}$-symmetry-breaking spin injection in the first biasing geometry, originating from the finite magnetization of a ferromagnet ($M \neq 0$), which generates a spin-polarized longitudinal current \textcolor{black}{and corresponding $\cal{T}$-odd spin current ($\textit{j}_\text{S}^{ODD}$)}. In the second biasing geometry, the spin injection \textcolor{black}{of $\textit{j}_\text{S}^{EVEN}$} is due to the $\cal{T}$-symmetric spin Hall effect as in the non-magnetic electrode. Bottom row: In this work, we demonstrate $\cal{T}$-symmetry-breaking spin injection from a compensated magnet ($M = 0$). \textcolor{black}{Unlike previously explored mechanisms, the spin injection is controlled by the magnetic state of the compensated magnet and is observed in both electrical biasing geometries.}}
\end{figure}

%RESULTS

\textcolor{black}{Our 15-nm-thick $\text{Mn}_5\text{Si}_3$ film, grown by molecular beam epitaxy (MBE) on a Si(111) substrate, has a critical temperature of T$_N \approx$ 240~K \cite{Reichlova2024}.} The high crystallinity of the epilayer was confirmed by X-ray diffraction (XRD) and transmission electron microscopy (TEM) \cite{Reichlova2024, kounta_competitive_2023} (see also Methods and Supplementary Fig.~S1). A 190-nm-wide injection electrode was patterned along the $a$-axis of $\text{Mn}_5\text{Si}_3$ (y-direction), as shown in Fig.~2a. A reference 120-nm-wide ferromagnetic $\text{Ni}_{0.8}\text{Fe}_{0.2}$ (permalloy, Py) injection electrode was deposited and patterned along y-direction on the same chip, together with 120-nm-wide ferromagnetic Py detection electrodes (running parallel to the injection electrodes) and a 120-nm-wide transverse non-magnetic Cu channel along x-direction. The edge-to-edge separation between injection and detection electrodes is 300 nm.

In Figs.~2b,c, we first check the reference LSV with the ferromagnetic Py injection electrode.  When a bias current ($\textit{I}_\text{bias}$) is applied from the injection electrode into the Cu channel (cf. left column of Fig. 1),  spin accumulation builds up at the Py/Cu interface, generating a pure spin current that diffuses through the Cu channel towards the Py detection electrode \cite{takahashi_spin_2003}. Here the resulting spin accumulation is observed via a non-local voltage ($\textit{V}_\text{NL}$) between the Py detection electrode and the Cu channel. In Fig.~2b, we plot the corresponding  non-local resistance, $\textit{R}_\text{NL}$ = $\textit{V}_\text{NL}$/$\textit{I}_\text{bias}$, showing the dependence on the relative magnetization orientations of the two ferromagnetic electrodes (black arrows in Fig.~2b), characteristic of the $\cal{T}$-symmetry-breaking spin injection from a ferromagnet \cite{johnson_interfacial_1985,jedema_electrical_2001,jedema_spin_2003,tombros_electronic_2007,lou_electrical_2007,casanova_control_2009,hamaya_estimation_2012,zahnd_spin_2018,fukuma_giant_2011,Kaiser2024}. The measurement of a spin signal, with the amplitude $\Delta\textit{R}_\text{NL}$ between parallel and antiparallel configuration, confirms that our Py electrodes and Cu channel work as expected, and also allows us to extract the in-plane coercive fields $\textit{B}_\text{c}$ $\approx$ 0.03 T and $\approx$ 0.1 T of the Py detection and injection electrodes, respectively, each one switching as a single magnetic domain at the junction with Cu, and with the magnetization pointing along the electrode length (y-direction). 

In Fig.~2c, we present reference measurements in the second biasing geometry corresponding to the top-right panel in Fig. 1, using the ferromagnetic injector and detector electrodes. The spin-Hall injection mechanism is detected when the magnetic field is applied along the x-direction \cite{omori_relation_2019}, \textcolor{black}{which ensures that the detector magnetization is oriented parallel or antiparallel to the polarization direction of the generated spin current (see the black arrows in Fig.~2c and Supplementary Fig.~S4).} The measured $\Delta\textit{R}_\text{NL}$ is only a fraction of that observed in the first biasing geometry, consistent with the small relativistic spin Hall angle ($\approx$ 1\%) \cite{omori_relation_2019} as compared to the large non-relativistic spin polarization (($\sigma^{\uparrow}-\sigma^{\downarrow}$)/($\sigma^{\uparrow}+\sigma^{\downarrow}$) $\approx$ \textcolor{black}{35}\%)\cite{sagastaSpinDiffusionLength2017a, dolanMeasuringMagneticAnisotropy2025} of the electrical current in Py, where $\sigma^{\uparrow(\downarrow)}$ correspond to the spin-dependent conductivities. As also expected for the spin-Hall mechanism, the measured signal in Fig. 2c is $\cal{T}$-symmetric with respect to the injector state - the measured $\textit{R}_\text{NL}$ signal only reflects changes in the magnetization of the Py detector and it saturates at $\textit{B}_\text{x}$ $\approx$ 0.2 T, which corresponds to the hard axis saturation field of the detection electrode. As a final consistency check of the reference LSV with Py injection electrode, we plot in Fig. 2c (inset) the measured signal using the second biasing geometry and a magnetic field applied along the y-direction. Here we confirm the expected absence of both the weak $\cal{T}$-symmetric spin injection signal (because the magnetization of the Py detector, along y-direction, is now perpendicular to the polarization direction of the spin current generated by the spin Hall effect) as well as the strong $\cal{T}$-symmetry-breaking spin injection. The small spurious signal in Fig. 2c (inset) arises from the non-ideal shape of the Py/Cu junctions. It is only a very small fraction of the strong signal observed in Fig. 2b.

The reference device with the ferromagnetic Py injection electrode serves two purposes. First, it confirms the state-of-the-art quality of design and nanofabrication of our LSVs. Second, the results obtained in the two biasing geometries of Py provide a direct benchmark for demonstrating the viability and qualitatively distinct phenomenology of the spin injection from the compensated magnet $\text{Mn}_5\text{Si}_3$. The side-by-side comparison between the two injection electrodes is shown in Figs. 2b,c and 2d,e. Figure 2d shows an analogous non-local signal to the one presented in Fig. 2b. The lower magnitude switching fields again correspond to reversing the magnetization of the Py detection electrode ($\textit{B}_\text{c}$ $\approx$ 0.03~T). The observed switching at higher fields, including intermediate steps, are due to  changes of the magnetic state of the $\text{Mn}_5\text{Si}_3$ injection electrode. Remarkably, we observe the spin signal in the first biasing geometry (Fig. 2d) despite the vanishing net magnetization of $\text{Mn}_5\text{Si}_3$ (see Supplementary Fig.~S2). Equally remarkably, the detected spin signal in the second biasing geometry, shown in Fig. 2e, is principally distinct from the one in Fig. 2c. Unlike the case of the reference measurements using the Py injection electrode (Figs.~2b and 2c), we observe similar patterns and amplitudes in the non-local resistances in both biasing geometries of the $\text{Mn}_5\text{Si}_3$ injection electrode (Figs.~2d and 2e). The similar results in Figs. 2d and 2e suggest a common mechanism of the spin injection from the compensated $\text{Mn}_5\text{Si}_3$ electrode in the two biasing geometries. The similarity of the measured non-local signals when using the reference ferromagnetic injection electrode in Fig. 2b and the compensated $\text{Mn}_5\text{Si}_3$ injection electrode in Figs.~2d and 2e hints that also in the latter case the spin injection is $\cal{T}$-symmetry-breaking. 

\begin{figure}[htbp]
	\centering
    \includegraphics[width=0.65\textwidth]{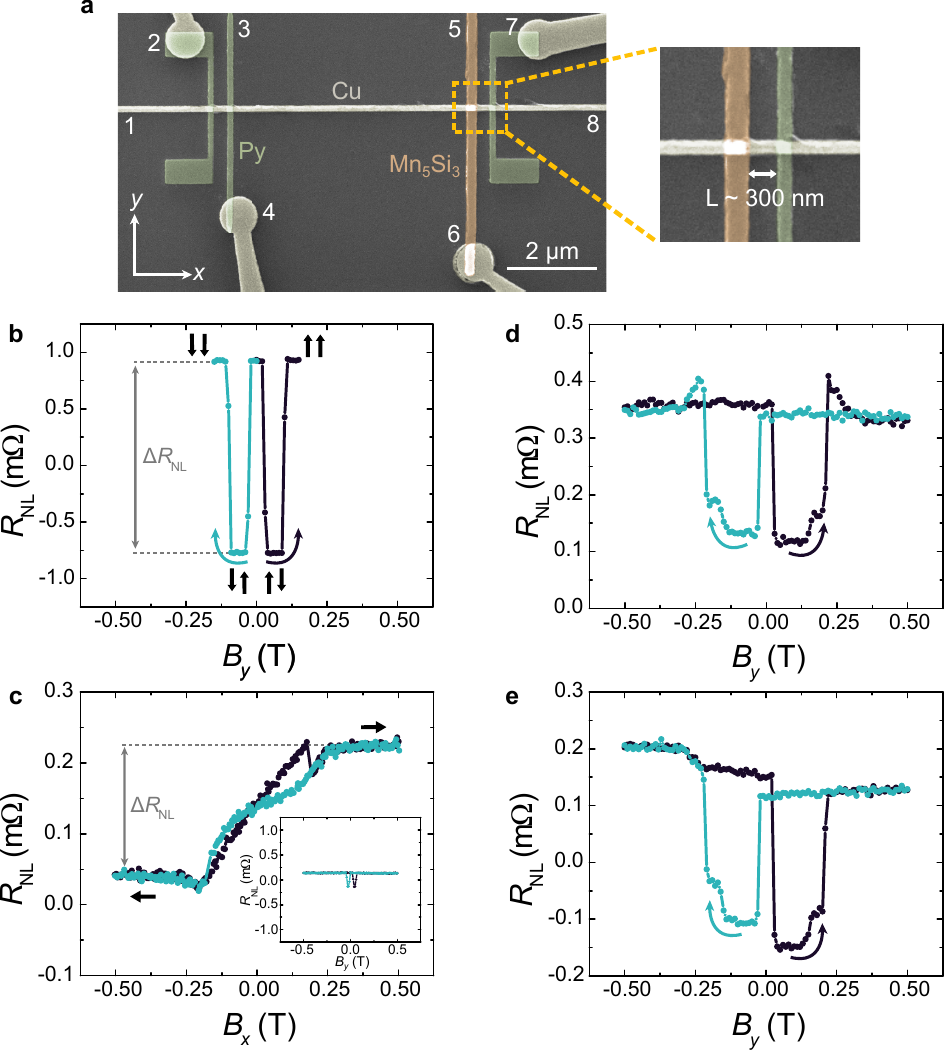}
	\caption{\textbf{Non-local spin valve measurements at 100 K.} \textbf{a} False-colored scanning electron microscope image of the Py/Cu/$\text{Mn}_5\text{Si}_3$ LSV device (right side of the main image) and a reference LSV device with Py injection and detection electrodes (left side of the main image). \textcolor{black}{The  edge-to-edge separation ($L=300$ nm) between the injection and detection electrodes is highlighted in the zoomed-in image.} \textbf{b} Non-local resistance $\textit{R}_\text{NL}$ (detected between contacts 1 and 2) for the reference Py/Cu/Py LSV in the biasing geometry across the Py/Cu interface (contacts 3 to 8). The magnetic field is along the y-direction. Black arrows show the magnetization direction of each Py electrode. Grey arrow shows the spin signal $\Delta\textit{R}_\text{NL}$. \textbf{c} $\textit{R}_\text{NL}$ (between contacts 1 and 2) for the reference Py/Cu/Py LSV in the biasing geometry along the Py electrode (contacts 3 to 4). The magnetic field is along the x-direction; inset: the same measurement with the field along the y-direction. Black arrows direction show the magnetization of the Py detector. Grey arrow shows the spin signal $\Delta\textit{R}_\text{NL}$. \textbf{d} $\textit{R}_\text{NL}$ for the Py/Cu/$\text{Mn}_5\text{Si}_3$ LSV (detected between contacts 7 and 8) in the biasing geometry across the $\text{Mn}_5\text{Si}_3$/Cu interface (contacts 5 to 1). The magnetic field is applied along the y-direction. \textbf{e} $\textit{R}_\text{NL}$ for the Py/Cu/$\text{Mn}_5\text{Si}_3$ LSV (detected between contacts 7 and 8) in the biasing geometry along the $\text{Mn}_5\text{Si}_3$ wire (contacts 5 to 6). The magnetic field is applied along the y-direction.} 
\end{figure}

To confirm the $\cal T$-symmetry-breaking nature of the spin injection from $\text{Mn}_5\text{Si}_3$, we fabricated a Hall cross in the $\text{Mn}_5\text{Si}_3$ injection electrode 3.5 $\upmu$m away from the junction with the Cu channel, as shown in Fig. 3a, to measure AHE. We verified that the $\text{Mn}_5\text{Si}_3$ electrode exhibits a temperature-dependent resistivity consistent with previous reports \cite{Reichlova2024, han_electrical_2024} (see Supplementary Fig.~S3a). As shown in Fig.~3b, we detect a remanent AHE (left axis), despite the vanishing remanent magnetization with an upper bound of 25 m$\upmu_\text{B}$/Mn (right axis and Supplementary Fig.~S2). Since the AHE is a $\cal T$-symmetry-breaking linear response \cite{nagaosa_anomalous_2010}, opposite signs of the measured remanent AHE evidence the switching between the time-reversed states of the compensated $\text{Mn}_5\text{Si}_3$ magnet. The two time-reversed remanent states are deterministically set by strong negative or positive magnetic fields applied along the in-plane y-direction. We point out that the AHE measurements were performed using the same in-plane field geometry as the non-local resistance measurements in the LSV. To connect our AHE measurements with previous studies of the AHE in $\text{Mn}_5\text{Si}_3$ thin films \cite{Reichlova2024,leiviska_anisotropy_2024, rial_altermagnetic_2024}, we show in Supplementary Fig.~S3b AHE measurements on our Hall cross with a magnetic field applied along the out-of-plane direction. Consistent with theoretical modeling \cite{rial_altermagnetic_2024}, we find that switching between the time-reversed states of $\text{Mn}_5\text{Si}_3$, accompanied by a reversal of the AHE sign, can be achieved using both out-of-plane and in-plane magnetic fields. \textcolor{black}{In addition, the AHE measurement in Fig.~3b indicates that the switching of the $\text{Mn}_5\text{Si}_3$ electrode proceeds through several intermediate steps. A possible origin of these states and their metastability are discussed in Supplementary Note V}. The non-local signal of the LSV measured in this device (Fig.~3c) exhibits a similar switching pattern, with intermediate steps and saturation above 0.5 T, providing additional evidence for the $\cal{T}$-symmetry-breaking nature of the spin current originating from $\text{Mn}_5\text{Si}_3$.

\begin{figure}[h!]
	\centering
    \includegraphics[width=0.8\textwidth]{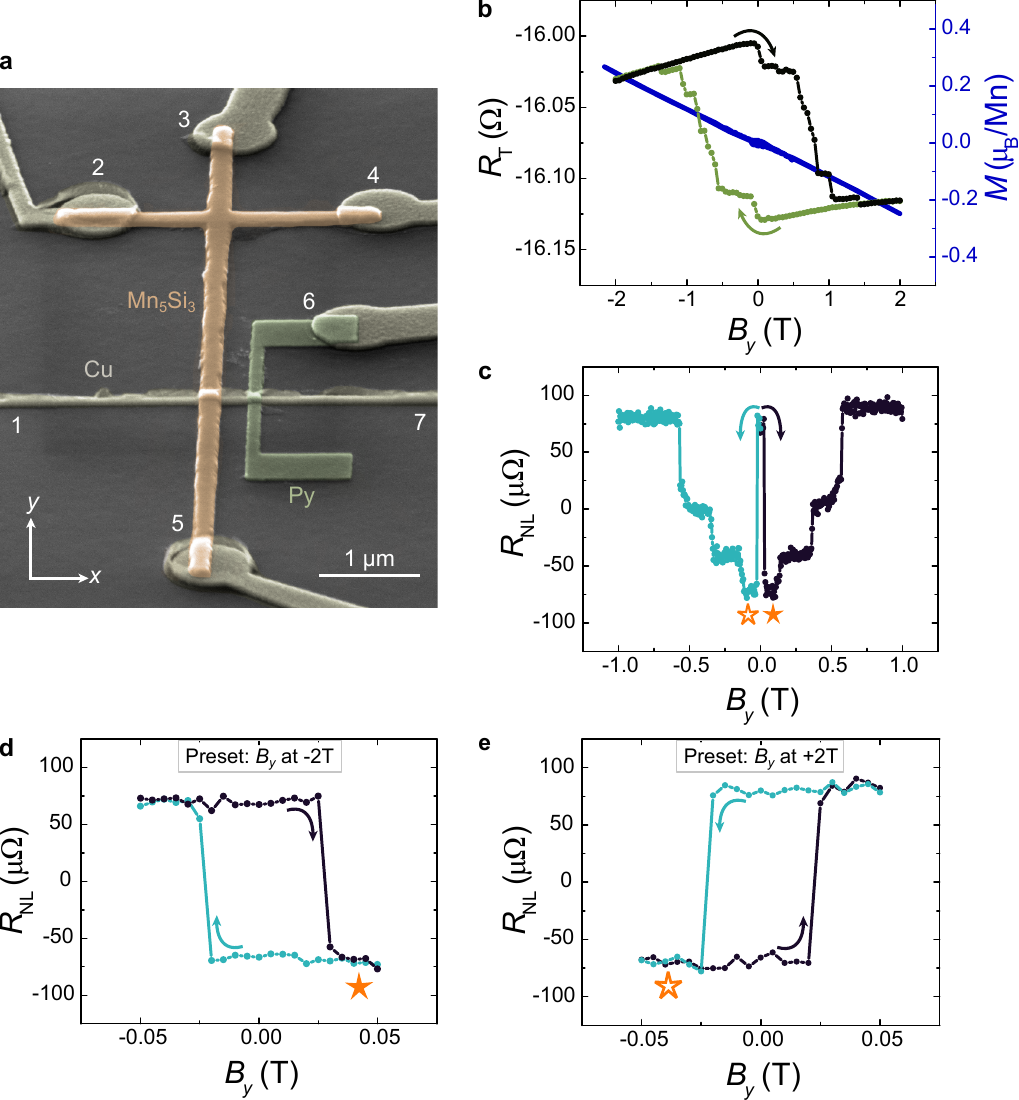}
	\caption{
    \textbf{$\cal T$-symmetry-breaking nature of the spin injection from $\text{Mn}_5\text{Si}_3$}. \textbf{a} False color SEM image of the device with defined Hall cross on the $\text{Mn}_5\text{Si}_3$ electrode. \textbf{b} Anomalous Hall effect measured in the $\text{Mn}_5\text{Si}_3$ wire (left axis) despite the vanishing net magnetic moment (right axis). The linear slope of the magnetization curve corresponds to the diamagnetic signal of the substrate (see Supplementary Fig.~S2 for details). The magnetic field is applied along the y-axis. \textbf{c} Non-local spin signal (detected between contacts 6 to 7) for the Py/Cu/$\text{Mn}_5\text{Si}_3$ LSV with biasing geometry across the $\text{Mn}_5\text{Si}_3$/Cu interface (contacts 3 to 1). The magnetic field is applied along the y-axis. The non-local signal exhibits intermediate steps as in the anomalous Hall signal. Stars mark states which are probed by measuring minor loops shown in panels \textbf{d} and \textbf{e} after presetting the system magnetization with -2 T and +2 T in-plane magnetic fields, respectively.}
\end{figure}

Figures~3d and 3e present further confirmation of the $\cal T$-symmetry-breaking spin injection from the $\text{Mn}_5\text{Si}_3$ electrode. As shown in the AHE measurements in Fig.~3b, the $\text{Mn}_5\text{Si}_3$ electrode can be deterministically switched between the two time-reversed states by applying a positive or negative strong magnetic field along the y-direction. In Fig.~3d, we put the $\text{Mn}_5\text{Si}_3$ electrode into a well-defined magnetic state by applying a $-2\,\mathrm{T}$ field, followed by ramping the field back to zero. Next, we applied a magnetic field of $-0.05\,\mathrm{T}$ to set the Py detection electrode magnetization along the $-$y direction. We then performed a field sweep from $-0.05\,\mathrm{T}$ to $+0.05\,\mathrm{T}$ and observe a switching of the non-local signal when the Py detector reverses its magnetization. Reversing the magnetic field sweep direction results in switching back the Py electrode at the negative coercive field, with the $\textit{R}_\text{NL}$ signal switching accordingly. To confirm the $\cal T$-symmetry-breaking nature of the injected spin current, in Fig.~3e we then set the $\text{Mn}_5\text{Si}_3$ electrode in the time-reversed magnetic state by applying a +2 T field, and then follow the same measurement protocol as in Fig.~3d. As expected, the sign of the hysteretic non-local signal in  Fig.~3e is opposite to that in Fig.~3d. The same time-reversed magnetic states can be prepared by applying an out-of-plane field as shown in Supplementary Fig.~S6. \textcolor{black}{Finally, we observe that the non-local signal vanishes at $\approx 200$~K, i.e., below $T_N$, again consistent with the $\cal T$-symmetry-breaking magnetic origin (see Supplementary Note~IX).}

%DISCUSSION

We now turn to the discussion of the microscopic origin for the experimentally observed $\cal T$-symmetry-breaking  spin injection from the compensated $\text{Mn}_5\text{Si}_3$ magnet. In the non-magnetic phase, the crystal unit cell of $\text{Mn}_5\text{Si}_3$ is hexagonal (space group $P6_3/mcm$), containing two formula units with four Mn atoms (Mn1) at Wyckoff position 4d and six Mn atoms (Mn2) at Wyckoff position 6g. Earlier neutron diffraction studies of bulk crystals have shown that, below the magnetic transition temperature, the magnetic unit cell becomes orthorhombic with two thirds of the Mn2 atoms carrying magnetic moments and forming a collinear antiferromagnetic phase with a doubling of the crystal unit cell along the \textit{b}-axis \cite{Gottschilch2012}. The resulting translation symmetry connecting opposite Mn2 magnetic moments in this antiferromagnetic phase excludes the AHE, in agreement with magnetotransport measurements in the $\text{Mn}_5\text{Si}_3$ bulk crystals (or thick polycrystalline films) \cite{Surgers2014,Surgers2016}, and in contrast to the remanent AHE seen in our $\text{Mn}_5\text{Si}_3$ thin films. The bulk antiferromagnetic phase also excludes the $\cal T$-symmetry-breaking spin injection found above in Figs. 2 and 3 in our thin films.

Recent experiments in $\text{Mn}_5\text{Si}_3$/Si(111) thin films have shown that, due to epitaxial constraints, the hexagonal unit cell is retained below the magnetic transition \cite{Reichlova2024,kounta_competitive_2023}. This, together with the lack of a magnetic frustration and the presence of the AHE and ANE \cite{Reichlova2024,leiviska_anisotropy_2024,Badura2025}, pointed towards altermagnetic ordering \cite{Smejkal2021a,Smejkal2022a,Reichlova2024} of the Mn2 atoms (Fig. 4a). The order is collinear and compensated but, unlike the antiferromagnetic order of bulk crystals, does not exhibit the doubling of the crystal unit cell. We note that a direct measurement of the magnetic order in the thin $\text{Mn}_5\text{Si}_3$ films is a major experimental challenge which goes beyond the scope of the work presented here (for more details see Supplementary Information of Ref. \cite{Reichlova2024}). Fig.~4a shows the opposite magnetic moments on Mn2 are not connected by translation or inversion but rather by mirror planes \textcolor{black}{(green dotted line)} and the state belongs to a d-wave altermagnetic class.

First-principles calculations of the non-relativistic $\cal T$-symmetry-breaking Fermi surface are shown in Fig.~4b. The d-wave altermagnetic state, apart from the AHE and ANE, allows for a $\cal T$-symmetry-breaking d-wave splitting of the non-relativistic Fermi surface into separate spin-up and spin-down channels \cite{Reichlova2024,leiviska_anisotropy_2024,Badura2025}. The preserved mirror symmetries \textcolor{black}{(green dotted line)} connecting the opposite magnetic moments on Mn2 in the altermagnetic phase protect the compensated nature of the magnetic ordering with vanishing net magnetization. The altermagnetic uniaxial distortion of the non-relativistic Fermi surfaces in each spin channel, and the mutually mirrored anisotropy axes between the two spin channels in d-wave altermagnetic $\mathrm{Mn}_5\mathrm{Si}_3$ (Fig.~4b), give rise to spin-dependent conductivity. As a result, electrical currents in the spin-up and spin-down channels, $\mathbf{j}^{\uparrow}$ and $\mathbf{j}^{\downarrow}$, can have different amplitudes and propagate at different angles with respect to the applied electric field \cite{smejkal_giant_2022,Gonzalez-Hernandez2021,Smejkal2021a,Smejkal2022a,jungwirth2026altermagnetic}. Specifically, when the electric field is applied along a spin-degenerate node of the d-wave Fermi surface, the resulting spin current $\mathbf{j}_s=\mathbf{j}^{\uparrow}-\mathbf{j}^{\downarrow}$ can emerge in the transverse direction, corresponding to the spin-splitter effect \cite{Gonzalez-Hernandez2021}. For electric-field directions away from the nodal axis, both longitudinal and transverse spin-current components are generated.

\begin{figure}[h!]
	\centering
    \includegraphics[width=0.9\textwidth]{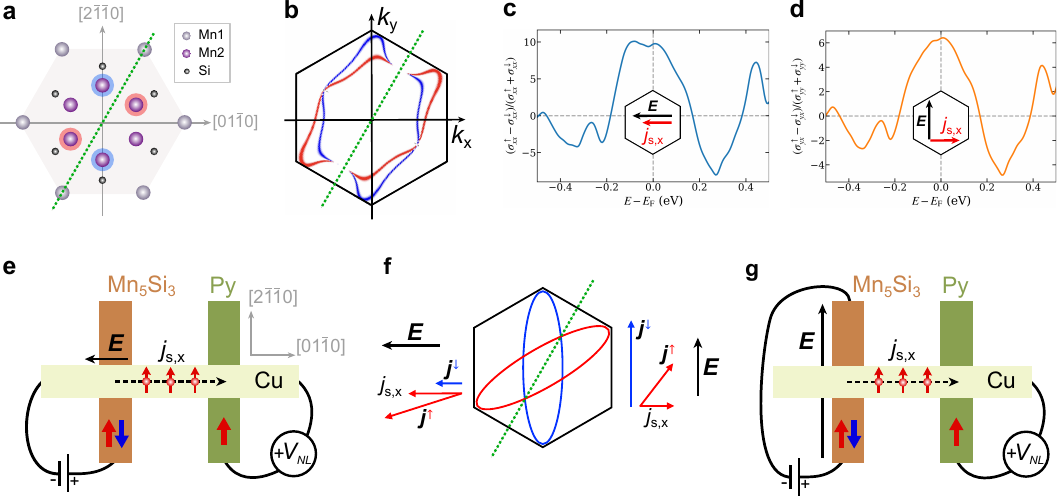}
	\caption{\textcolor{black}{\textbf{Spin-dependent transport in the altermagnetic phase of $\text{Mn}_5\text{Si}_3$.} (\textbf{a}) Magnetic and crystal structure of the altermagnetic phase of $\text{Mn}_5\text{Si}_3$ and (\textbf{b}) the corresponding calculated spin-resolved Fermi surfaces, with spin-down (blue) and spin-up (red) states and the mirror planes indicated by green dotted lines. The calculated spin polarization for the two experimental electrical-biasing geometries: (\textbf{c}) longitudinal spin polarization $(\sigma_{xx}^{\uparrow}-\sigma_{xx}^{\downarrow})/(\sigma_{xx}^{\uparrow}+\sigma_{xx}^{\downarrow})$ and (\textbf{d}) transverse spin polarization $(\sigma_{yx}^{\uparrow}-\sigma_{yx}^{\downarrow})/(\sigma_{yy}^{\uparrow}+\sigma_{yy}^{\downarrow})$, where $\sigma_{ij}^{\uparrow / \downarrow}$ are elements of the spin-dependent conductivity tensor. The electric field direction is indicated in the panel insets. The schematic spin-up and spin-down Fermi surfaces in panel \textbf{f} illustrate how a finite spin-current component can be injected into the Cu channel in both experimental biasing geometries. For the geometry shown in panel \textbf{e}, the injected spin current originates predominantly from the longitudinal component of the spin current, whereas for the geometry shown in panel \textbf{g} it originates from the transverse component.} 
    }
\end{figure}

\textcolor{black}{In our LSV devices, spin injection into the Cu channel is determined solely by the $x$-component of the spin current. Depending on the biasing geometry, this injected spin current originates predominantly from either the longitudinal or the transverse component of $\mathbf{j}_s$. This situation is illustrated schematically in Figs.~4e and 4g for the two experimental biasing geometries, considering the crystallographic orientation of the $\mathrm{Mn}_5\mathrm{Si}_3$ injector patterned along the $a$-axis ([2$\bar{1}\bar{1}$0]). The schematics of Fig.~4f visualize the corresponding spin-up and spin-down transport channels and show how their combination generates a finite spin-current component along $x$ in both geometries.}

\textcolor{black}{The magnitude of the injected spin current is quantified by the spin polarization of the dissipative symmetric longitudinal and transverse components of the conductivity tensor in the conserved spin-up and spin-down transport channels \cite{Reichlova2024}, shown in Figs.~4c,d. The calculated spin polarization of the conductivity reaches approximately 10$\%$ and 6.5$\%$ for the two biasing geometries, respectively, indicating that comparable spin injection is expected despite the different directions of the applied bias. These sizable $\cal T$-odd spin currents are generated by the non-relativistic altermagnetic ordering, are intrinsic to the material, and are essentially unaffected by quasiparticle scattering. Furthermore, they are only weakly modified by spin-orbit coupling in $\mathrm{Mn}_5\mathrm{Si}_3$. The weak spin-orbit coupling is also reflected in the order-of-magnitude smaller $\cal T$-even spin Hall contribution compared to the $\cal T$-odd altermagnetic mechanism (see Supplementary Note XII).}

\textcolor{black}{Upon switching between the two time-reversed altermagnetic states, the calculated spin polarization reverses sign, leading to an opposite injected spin current. This prediction is consistent with the measurements in Figs.~2d,e, where the two biasing geometries produce non-local spin signals of comparable magnitude and opposite sign for opposite time-reversed states. Overall, the agreement between the first-principles calculations and the non-local spin valve measurements supports a picture of conserved spin-up and spin-down transport channels generated by the non-relativistic altermagnetic ordering in $\mathrm{Mn}_5\mathrm{Si}_3$.}

By symmetry, the AHE is not allowed in the altermagnetic phase of $\text{Mn}_5\text{Si}_3$ for spins aligned along the out-of-plane $c$-axis. The observed remanent AHE in Fig.~3b is thus consistent with the presence of an in-plane component of the magnetic easy axis in the altermagnetic phase of $\text{Mn}_5\text{Si}_3$. The polarization axis of the injected spin current, resulting from the non-relativistic spin polarization of the Fermi surfaces, is aligned with the altermagnetic order parameter and therefore also contains an in-plane component. The in-plane altermagnetic origin of the spin injection is thus also consistent with the non-local signals measured using the in-plane magnetized Py detection electrode in Figs.~2 and 3.

Finally, we highlight that the electrical current flowing along the out-of-plane [0001] crystallographic direction of $\text{Mn}_5\text{Si}_3$ is not spin-polarized in the altermagnetic phase. This implies that only the in-plane current is spin-polarized and contributes to spin injection, i.e., spins are injected through the vertical side-edge interface between the $\text{Mn}_5\text{Si}_3$ electrode and the Cu channel. We consistently  observe a sizable spin signal only in devices subjected to a careful etching treatment of the vertical side edges of the $\text{Mn}_5\text{Si}_3$ electrode (see Supplementary Fig.~S10). In contrast, spin injection from the reference Py electrode occurs predominantly through the horizontal top interface with the Cu channel, rendering the quality of the side-edge interface relatively unimportant. 

\textcolor{black}{To further confirm the spin injection through the side edges, we fabricated a control device in which transport through the top surface of the $\text{Mn}_5\text{Si}_3$ electrode to the Cu channel is blocked by a 12-nm-thick SiO$_2$ interfacial  barrier. Supplementary Fig.~S11 shows that the sizable non-local signal persists in this control sample in both biasing geometries (for more details see Supplementary Note VIII). We note that the non-local signal observed in Supplementary Fig.~S11 suggests bistable switching of the $\text{Mn}_5\text{Si}_3$ electrode in this device. The presence or absence of intermediate states may reflect either the complex energy landscape of uniform $\text{Mn}_5\text{Si}_3$ or the formation of non-uniform magnetic-domain states, depending on the orientation of the altermagnetic order parameter and the applied magnetic field. In both scenarios, the occurrence of intermediate states can be sensitive to sample-to-sample variations in strain, defects, and disorder (for more details, see Supplementary Note V). While the detailed switching pathway may therefore vary between devices, the resulting non-local signal is robust and reproducible. As shown in Supplementary Note~VI, measurements over multiple magnetic-field cycles exhibit negligible variations in the detected signal. Finally, in Supplementary Note X, we present measurements on the device with the SiO$_2$ top interfacial barrier for an in-plane field applied along the hard axis ($x$-axis) of the Py detector, and compare them with measurements performed with the field along the easy axis ($y$-axis). Note that, in the reference LSV with a Py injection electrode, the measurement with the magnetic field applied along the $x$-axis detects the $\cal T$-symmetric spin Hall injection through the top Py/Cu interface. In contrast, this vertical injection pathway is blocked in the device with the $\text{Mn}_5\text{Si}_3$ injector by the insulating SiO$_2$ barrier separating the injector from the Cu channel. Nevertheless, a sizable non-local signal persists even for the magnetic field applied along $x$. The persistence of the signal provides further evidence that spin injection originates predominantly from the Mn$_5$Si$_3$/Cu side-edge interface and is dominated by the $\cal T$-symmetry-breaking component.}

\textcolor{black}{LSVs are not competitive to vertical nanoscale junctions for spintronic applications due to low output voltage signals and poor scalability. They are, however, invaluable for disentangling the fundamental spintronic processes of injection, propagation and detection of spin currents, and for determining basic physical parameters governing these processes. Spin diffusion length is one of these key parameters. By combining our experimental results and the calculated spin polarization with quantitative modeling, we determine the spin diffusion length in $\text{Mn}_5\text{Si}_3$. We model the device using the established one-dimensional spin diffusion formalism for LSVs, extended to incorporate an altermagnetic $\text{Mn}_5\text{Si}_3$ injector, as detailed in Supplementary Note XI. All relevant device parameters, including the spin diffusion length of Cu, the spin transport properties of the Py detector, the device dimensions, resistivities, and interface resistances, are independently determined, leaving only the spin polarization and spin diffusion length of $\text{Mn}_5\text{Si}_3$ as free parameters. By considering the spin polarization of $\approx 6.5-10\%$, obtained from our first-principles calculations (Figs. 4c,d), we obtain a spin diffusion length of approximately $50-130$~nm for $\text{Mn}_5\text{Si}_3$. Notably, this value is at least an order of magnitude larger than the typical spin diffusion lengths in Py or other common ferromagnets \cite{sagastaSpinDiffusionLength2017a, dolanMeasuringMagneticAnisotropy2025}. The long spin diffusion length, consistent with the collinear compensated altermagnetic order and the weak spin-orbit coupling in $\text{Mn}_5\text{Si}_3$, is another distinctive merit in comparison to the conventional ferromagnetic and spin Hall materials.}

\textcolor{black}{To conclude, our results establish a proof-of-concept demonstration that altermagnets can act as longitudinal and transverse spin-current sources controlled by the magnetic state, without requiring magnetization, non-collinear magnetic ordering or strong spin-orbit coupling. }\textcolor{black}{While only $d$-wave altermagnets allow for the non-relativistic spin-dependent conductivities \cite{Gonzalez-Hernandez2021,dou2025anisotropic}, we note that $g$-wave and $i$-wave altermagnets can also host spin-dependent transport phenomena in tunneling geometries \cite{smejkal_giant_2022}, which greatly expands the landscape of potentially suitable altermagnetic materials for spintronics research and applications.}

\section{Methods}

\subsection{Epitaxial crystal growth}

The 18-nm-thick $\text{Mn}_5\text{Si}_3$ thin films were deposited using MBE within an ultra-high vacuum chamber, with a base pressure below 9 x $10^{-10}$ mbar. The deposition was carried out on an intrinsic Si(111) substrate with a resistivity exceeding 10,000 $\Omega$·cm. Manganese and silicon were co-evaporated from a Knudsen cell and a sublimation cell, respectively. The fluxes were precisely controlled to achieve the desired stoichiometry. In situ reflection high-energy electron diffraction (RHEED) was employed to monitor the growth process, revealing the distinctive ($\sqrt{3}$ × $\sqrt{3}$)R30° reconstruction characteristic of the $\text{Mn}_5\text{Si}_3$ phase during post-annealing up to 300°C.

\subsection{Transmision electron microscopy and X-ray diffraction}

The crystal structure of the as-grown films was examined ex situ using XRD with a high-brilliance rotating anode and TEM on a Jeol JEM-2100F microscope operating at 200 kV, providing a spatial resolution of 2.3 \textup{~\AA}. Cross-sectional TEM samples were prepared using a focused dual ion beam (FEI Helios 600 NanoLab). Both XRD and TEM analyses confirmed the good crystalline quality of the $\text{Mn}_5\text{Si}_3$ thin films whose growth was assisted by a thin MnSi seed layer. The epitaxial relationships were determined to be: $\text{Mn}_5\text{Si}_3$ (0001)-[010] $\parallel$ Si(111)-[10]. For a more comprehensive description of the growth process, refer to Ref.\cite{kounta_competitive_2023} and Supplementary Fig.S1.

\subsection{Magnetometry measurements}

Magnetization measurements were performed using a Quantum Design MPMS3 system in VSM mode with a standard quartz holder. The same wafer used for transport measurements was cut into a 3.5 × 5 mm piece and measured in the same experimental geometry as the LSVs, with the magnetic field applied in the y-direction. The magnetometry signal is dominated by the 0.5-mm-thick silicon substrate, which contributes to a strong diamagnetic background. To estimate the upper limit of the net magnetic moment in our samples, we subtracted the linear diamagnetic background and plotted the residual signal in the inset of the figure (see Supplementary Fig. S2). The diamagnetic contribution near zero external magnetic field is negligible, allowing us to estimate the spontaneous net magnetic moment arising from the films to be at most 25 m$\upmu_\text{B}$/Mn, as indicated by the orange background. This value is consistent with previous reports on $\text{Mn}_5\text{Si}_3$.

\subsection{Sample fabrication}

We carried out nanofabrication on epitaxial $\text{Mn}_5\text{Si}_3$ thin films to define the $\text{Mn}_5\text{Si}_3$ electrodes. A 5-nm-thick $\text{SiO}_2$ layer was RF-sputtered to protect the film, followed by spin-coating with AR-N 7520.17 negative e-beam resist. The electrode, which includes a Hall cross, was patterned by e-beam lithography and developed in AR 300-47. Exposed areas were etched by Ar-ion milling before removing the residual resist in acetone. Next, the sample was coated with ZEP 520 A7 positive e-beam resist to define the ferromagnetic electrodes via e-beam lithography. After development in ZED-N50, we deposited 30 nm of Py by e-beam evaporation in ultra-high vacuum ($3\times10^{-9}$ mbar), and the residual resist was removed using 1-methyl-2-pyrrolidone (NMP). We used Ar-ion milling to clean the electrode side edges with the Ar-ion beam directed nearly parallel to the sample surface. In the following step, the sample was coated with a double layer of polymethyl methacrylate (PMMA) positive e-beam resist to pattern the spin channel by e-beam lithography. After development in methyl isobutyl ketone/isopropanol (MIBK/IPA 1:3) solution, the electrode interfaces were cleaned by Ar-ion milling at a near-perpendicular angle. A 70-nm-thick Cu layer was deposited by thermal evaporation in ultra-high vacuum ($5\times10^{-9}$ mbar), and the resist was removed in acetone. Finally, the nanostructures were capped with a 5-nm-thick SiO$_2$ layer by RF-sputtering to prevent surface oxidation.

\subsection{Magnetotransport measurements}

Magnetotransport measurements were carried out using a Physical Property Measurement System (PPMS) developed by Quantum Design. The system is equipped with a liquid helium cryostat, enabling temperature control in the range of 1.9 K to 400 K, and a superconducting magnet capable of generating magnetic fields up to ±9 T. The sample was mounted on a rotating holder to allow variation of its orientation with respect to the magnetic field. Electrical current was supplied using a Keithley 6221 current source, and voltage was measured with a Keithley 2182A nanovoltmeter in the various measurement geometries used (4-point resistance, AHE and non-local resistance as shown in Fig. 1). Measurements were performed in delta mode, averaging the measured voltages obtained for alternating current polarities to eliminate thermoelectric effects and reduce the noise.

\subsection{Electronic band structure and spin current calculations}

Ab initio calculations were performed within the density functional theory framework as implemented in the Vienna Ab Initio Simulation Package \cite{Kresse1996a}. The hexagonal unit cell with the lattice parameters $a = b = 6.901$\,\AA\ and $c = 4.795$\,~\AA~was used, consistent with experimental reports\cite{Gottschilch2012,Reichlova2024}. The Perdew–Burke–Ernzerhof (PBE) exchange–correlation functional \cite{Perdew1996} was employed to describe the electronic interactions within the generalized gradient approximation. A plane-wave cutoff energy of $500\,\mathrm{eV}$ and a $\Gamma$-centered $9\times9\times9$ k-point mesh were used for Brillouin zone sampling. Spin-polarized collinear calculations were performed. The maximally localized Wannier functions (MLWFs) were constructed using the Wannier90 package \cite{Pizzi2020}, with Mn-s,d and Si-s,p orbitals selected as initial projectors. 

The Fermi surfaces were obtained using WannierTools \cite{Wu2017b}, based on the spin-resolved Wannier Hamiltonian. Separate calculations were performed for the spin-up and spin-down channels on a $480\times480\times1$ k-mesh in the $k_z=0$ plane, enabling visualization of the spin-polarized Fermi surface. The Boltzmann transport properties were evaluated using the BoltzWann module of Wannier90 within the constant relaxation time approximation (using spectral broading $\Gamma=0.01\,\mathrm{eV}$). The electrical conductivity tensor was computed using a $160\times160\times160$ k-point grid.

\clearpage
\medskip
\textbf{Data Availability Statement}\par
The data that supports the findings of this study are available upon reasonable request.

\medskip
\textbf{Acknowledgements} \par 

The authors acknowledge funding from MICIU/AEI/10.13039/501100011033 (Grants No. CEX2020-001038-M and CEX2025-001634-M), from MICIU/AEI and ERDF/EU (Project No. PID2024-155708OB-I00), and from the European Union’s Horizon 2020 research and innovation programme under the Marie Skłodowska-Curie Grant Agreement No. 955671 (SPEAR).  We also acknowledge Grant Agency of the Czech Republic Grant No. 22-17899K, The MEYS of the Czech Republic through the OP JAK call Excellent Research (TERAFIT Project No. CZ.02.01.01/00/22\_008/0004594), the Dioscuri Program LV23025 funded by Max Planck Society and MEYS of the Czech Republic, German Federal Ministry of Education and Research, COST action CHIROMAG. This work was also supported by the French National Research Agency (ANR) under the PEPR SPIN project ALTEROSPIN (grant number ANR-24-EXSP-0002). J.M. acknowledges funding by the Department of Education of the Basque Government under the Predoctoral Programme for the training of nondoctoral research staff (Grant No. PRE\_2022\_1\_0297). L.S. acknowledges funding from the ERC Starting Grant No. 101165122. D.K. acknowledges the Lumina Quaeruntur fellowship LQ100102201 of the Czech Academy of Sciences. O.G. and J.S. acknowledge support of DFG (Elasto-Q-Mat, TRR 288 – 422213477 projects A12 and A09, and Spin+X TRR 173-268565370 projects A03 and B15). N.S. acknowledges support from the Programa de Desarrollo de las Ciencias Básicas (PEDECIBA). T. K. acknowledges support by the Research Council of Finland through DYNCOR, Project Number 354735 and through the Finnish Quantum Flagship, Project Number 359240.  Moreover, his work is part of the Finnish Centre of Excellence in Quantum Materials (QMAT). F.S.B. thanks financial support from the Spanish MICIU/AEI/10.13039/501100011033 through Project No. PID2023-148225NB-C31,  the European Union’s Horizon Europe research and innovation program under grant agreement No. 101130224 (JOSEPHINE), and the Finish InstituteQ under the Chair of Excellence programme.

\medskip
\textbf{Author Contributions}\par
H.R. and F.C. conceived the research. I.K., M.P., C.G., A.B. and L.M prepared and characterized $\text{Mn}_5\text{Si}_3$ thin films. J.M. fabricated the samples, with input from A.B., S.B, and J.R. J.M. performed the electrical measurements and the data were analyzed by J.M., O.G., T.J., H.R., and F.C. \textcolor{black}{E.D. performed the finite element method simulations.} A.B.H., W.C., \textcolor{black}{R.G.-H.}, and L.S. performed the first-principles calculations. \textcolor{black}{N.S., T.K., and F.S.B. developed the quantitative description of the spin signal based on the spin diffusion formalism.} D.K., J.S., V.B., and L.E.H. contributed to discussions and data analysis. J.M., F.C., T.J., and H.R. co-wrote the manuscript with input from all authors. All authors discussed the results and commented on the manuscript.

\medskip
\textbf{Competing Interest}\par
The authors declare no competing interests.

%\bibliography{Main/refsTJ, Main/refsHelena, Main/refsSebas, SI/refsHelena2}
%\bibliography{filtered_references_final_4}

%\printbibliography[]
%\bibliographystyle{plain}
%\bibliographystyle{unsrtnat}
\end{document}